\documentclass{article}
\usepackage{amsmath}
\usepackage{amsthm} 
\usepackage{amssymb}
	
\usepackage{bussproofs}
\usepackage{url,xspace}
\usepackage{mathbbol}
\usepackage[all]{xy}
\SelectTips{cm}{}

\theoremstyle{plain} % default

\theoremstyle{definition}

\newcommand{\pure}{{(0)}} % pure
\newcommand{\Exc}{\mathit{Exc}}
\newcommand{\ppg}{{(1)}} % propagator or thrower  
\newcommand{\ctc}{{(2)}} % catcher 
\newcommand{\St}{\mathit{St}}
\newcommand{\acc}{{(1)}} % accessor or observer
\newcommand{\modi}{{(2)}} % modifier 
\newcommand{\eqs}{\equiv} % strong 
\newcommand{\eqw}{\sim} % weak 
\newcommand{\inl}{\mathit{in}_1} % left coprojection 
\newcommand{\prl}{\mathit{pr}_1} % left projection 
\newcommand{\tuple}[1]{(#1)}
\newcommand{\deno}[1]{[[#1]]}

\title{Scalability using effects} 
\author{Dominique Duval
\thanks{Universit\'e de Grenoble, Laboratoire Jean Kuntzmann, 
\texttt{Dominique.Duval@imag.fr}.}}
\date{10 June 2013 --- SLS 2013 --- Extended abstract}
 
\begin{document}

%------------------------------------------------------------------------- 
\maketitle

\begin{abstract}
This note is about using computational effects 
for scalability. With this method, 
the specification gets more and more complex 
while its semantics gets more and more correct. 
We show, from two fundamental examples, 
that it is possible to design a deduction system for 
a specification involving an effect without expliciting this effect.
\end{abstract}

%------------------------------------------------------------------------- 
\section{Introduction}

A well-known pedagogical trick for teaching complex features 
is to ``lure'' the students by first providing a simplified version
of this feature, before adding the required corrections to this 
approximate version. Typically:
``\textit{The plural form of most nouns is created by adding the letter 's' 
to the end of the word, but there are some exceptions\dots}''. 
In computer science, such an approach is used in the mechanism of
\emph{exceptions}, as well as in many other \emph{computational effects}.
The aim of this note is to present computational effects
as a tool for solving some scalability issues 
in the design of specifications. 

There are several approaches for managing a large specification (or program).
For instance, \emph{components} and \emph{modules} can be used 
for breaking down the specification into smaller pieces; 
the intended semantics of the whole specification is obtained by 
``merging'' the semantics of its components. 
Then each module is ``right'', in the sense that 
its semantics describes some ``part'' of the intended semantics. 
Another method is the \emph{stepwise refinement}, 
where one builds progressively 
the required specification from more abstract specifications; 
during the refinement process, 
the specification gets more and more complex 
while its semantics gets more and more precise, 
until the intended semantics is obtained. 
Then each step is ``right'', in the sense that the intended semantics is
one of the possible semantics of each intermediate specification. 

In this note we focus on using \emph{computational effects} 
for scalability. With this method, 
the specification gets more and more complex 
while its semantics gets more and more \emph{correct}. 
Thus, each step is ``wrong'', in the sense that the intended semantics is
not a semantics of any intermediate specification. 
It follows that the main issue for using computational effects 
for scalability is whether it is possible 
to use the intermediate specifications for testing or verifying or proving 
non-trivial properties of the final specification. 
In the next sections we argue in favour of a positive answer to this question, 
from two examples which both rely on a common general algebraic framework.
The key point is that each intermediate specification may become 
``right'' simply by classifying its features according to the 
way they behave with respect to the corresponding effect,
without describing explicitly this behaviour. 

This note is based on work with J.-C.~Reynaud, J.-G.~Dumas, L.~Fousse
and C.~Dom\'inguez. 

%------------------------------------------------------------------------- 
\section{Exceptions} 
\label{exc}
 
The mechanism of \emph{exceptions} has two parts: 
\emph{raising} exceptions (with keywords \texttt{throw} or \texttt{raise})
and \emph{handling} them (with keywords \texttt{try/catch} or \texttt{handle}). 

The lure in the raising of exceptions lies in the fact 
that a function $f$ which takes an argument of type $A$ and returns a value 
of type $B$ is not interpreted as a map from $\deno{A}$ to $\deno{B}$ 
(where $\deno{T}$ denotes the interpretation of a type $T$) 
but as a map $\deno{f}:\deno{A}\to\deno{B}+\Exc$, 
where $\Exc$ is the set of exceptions and ``$+$'' denotes the disjoint union. 
This can be expressed in a categorical framework \cite{Mo91}: 
a function $f:A\to B$ is interpreted as a map $\deno{f}:\deno{A}\to \deno{B}$ 
in the Keisli category of the monad $X\mapsto X+\Exc$ on the category of sets. 
The lure in the handling of exceptions is that  
a function $f:A\to B$ inside a \texttt{catch} clause may recover 
from an exception (it may also raise an exception), 
so that is interpreted as a map $\deno{f}:\deno{A}+\Exc\to\deno{B}+\Exc$.

Thus, in order to deal with exceptions, our proposal is to add 
\emph{decorations} to the syntax: 
a \emph{catcher} $f^\ctc:A\to B$ is interpreted as 
$\deno{f}:\deno{A}+\Exc\to\deno{B}+\Exc$
while a \emph{propagator} $f^\ppg:A\to B$ is interpreted as 
$\deno{f}:\deno{A}\to\deno{B}+\Exc$ 
and a \emph{pure} function $f^\pure:A\to B$ is interpreted simply as 
$\deno{f}:\deno{A}\to\deno{B}$.
In order to prove properties of programs involving exceptions, 
we must also add decorations to the equations: 
when $f,g:A\to B$ are catchers (which is the general case), 
a \emph{strong} equation $f\eqs g$ means that 
$\deno{f}=\deno{g}:\deno{A}+\Exc\to\deno{B}+\Exc$
and a \emph{weak} equation $f\eqw g$ means that 
$\deno{f}\circ\inl =\deno{g}\circ\inl :\deno{A}\to\deno{B}+\Exc$,
where $\inl:\deno{A}\to\deno{A}+\Exc$ is the left coprojection. 

A major point is the existence of a \emph{deduction system}
associated to these decorations, 
which can be used for proving properties of specifications
using exceptions. 
For instance, there are rules for the obvious hierarchies: 
each pure function can be seen as a propagator, 
each propagator as a catcher, 
and each strong equation as a weak one. 
There are also rules expressing the facts that 
composition preserves the decorations, 
that the strong equations generate a congruence, 
and that the weak equations generate a kind of ``weak'' congruence: 
if $f_1\eqw f_2$ then $f_1\circ g\eqw f_2\circ g$ (but in general 
$h\circ f_1 \not\eqw h\circ f_2$). 
More details can be found in \cite{DDFR12-exc}. 

%------------------------------------------------------------------------- 
\section{States}
\label{state}

Imperative programming relies on the notion of side-effect 
for states: 
a state of the memory can be \emph{observed} thanks to 
\texttt{lookup} functions 
and it can be \emph{modified} thanks to 
\texttt{update} functions. 

The lure in this situation is 
that a function $f$ which takes an argument of type $A$ and returns a value 
of type $B$ may be interpreted either as 
$\deno{f}:\deno{A}\times\St\to\deno{B}$ if $f$ is an observer
or as $\deno{f}:\deno{A}\times\St\to\deno{B}\times\St$ 
if $f$ is a modifier (where $\St$ is the set of states 
and ``$\times$'' the cartesian product).  
In order to deal with imperative programs, we add 
\emph{decorations} to the syntax: 
a \emph{modifier} $f^\modi:A\to B$ is interpreted as 
$\deno{f}:\deno{A}\times\St\to\deno{B}\times\St$
while an \emph{observer} $f^\acc:A\to B$ is interpreted as 
$\deno{f}:\deno{A}\times\St\to\deno{B}$.
As with exceptions, a \emph{pure} function $f^\pure:A\to B$ 
is interpreted simply as $\deno{f}:\deno{A}\to\deno{B}$.
We also add decorations to the equations: 
when $f,g:A\to B$ are modifiers (this is the general case), 
a \emph{strong} equation $f\eqs g$ means that 
$\deno{f}=\deno{g}:\deno{A}\times\St\to\deno{B}\times\St$
and a \emph{weak} equation $f\eqw g$ means that 
$\prl\circ\deno{f} =\prl\circ\deno{g} :\deno{A}\times\St\to\deno{B}$ 
where $\prl:\deno{B}\times\St\to\deno{B}$ is the left projection. 
This can also be expressed in a categorical framework: 
an observer $f^\acc:A\to B$ is interpreted as a map 
$\deno{f}:\deno{A}\to \deno{B}$ 
in the co-Keisli category of the comonad $X\mapsto X\times\St$ 
on the category of sets,
while a modifier $f^\modi:A\to B$ may be interpreted as a map 
$\deno{f}:\deno{A}\to \deno{B}$ 
in the Keisli category of the monad $X\mapsto (X\times\St)^\St$ \cite{Mo91}. 
For instance, 
when dealing with a toy class for bank accounts 
in an object oriented language, the expressions  
  ``$ \texttt{deposit(7); balance()}; $''
and 
  ``$\texttt{7 + balance()}; $'' 
have different effects but they return the same integer.  
They can be seen as decorated terms
  $ f^\texttt{(2)}=
  \texttt{balance}^\texttt{(1)}\circ \texttt{deposit}^\texttt{(2)}\circ 
  \texttt{7}^\texttt{(0)} $
and 
  $ g^\texttt{(1)}=\texttt{+}^\texttt{(0)}\circ 
  \tuple{\texttt{7}^\texttt{(0)},\texttt{balance}^\texttt{(1)}} $, 
and indeed it can be proved that $f^\texttt{(2)}\eqw g^\texttt{(1)}$
without introducing any type of states.

In fact, our approach reveals a duality between exceptions and states 
\cite{DDFR12-dual}. Thus, we get rules for 
proving properties of imperative programs which are dual 
to the rules mentioned above for exceptions. 
More details are given in \cite{DDFR12-state}. 
In this duality, the exception monad $X+E$ corresponds to the comonad 
$X\times\St$, but there is no comonad on sets corresponding to the 
states monad $(X\times\St)^\St$; from our point of view 
this is not an issue, in contrast with the point of view of monads and
Lawvere theories where there is a need for specific tools for 
\emph{handlers} \cite{PP02,PP09}. 

%------------------------------------------------------------------------- 
\section{Conclusion} 

In the previous sections we have outlined the construction of 
deduction systems for dealing with exceptions and states 
in an \emph{implicitly} way, i.e., 
without using any ``type of exceptions'' or ``type of states'';
we have simply added some decorations to the syntax of the language. 
This can be done for other computational effects:  
several examples are given in \cite{DDR11-seqprod}, 
where we propose a formalization of 
the fact that the order of evaluation of the arguments 
of a binary function becomes crucial in presence of effects. 

This work relies on categorical tools \cite{DD10-dialog}:  
mainly on \emph{adjunction},
more precisely on \emph{categories of fractions} \cite{GZ67},
and on \emph{limit sketches} \cite{Eh68,BW99}. 

We have described one step in our approach to scalability. 
Thanks to the categorical framework it should be easy to express 
the composition of effects, thus getting a stepwise scalability method. 
The combination of this method with other scalability methods 
still has to be studied.  

%------------------------------------------------------------------------- 
%------------------------------------------------------------------------- 

%------------------------------------------------------------------------- 
%------------------------------------------------------------------------- 
\end{document}